\documentclass[12pt]{article}
\usepackage{epsfig}
\usepackage{wrapfig}
\usepackage{subfigure}
\usepackage{amsmath}
\usepackage{hhline}
\usepackage{amssymb}
\usepackage{times}
\usepackage{cite}

\usepackage[]{lineno}

\newcommand{\be}{\begin{equation}}
\newcommand{\ee}{\end{equation}}
\newcommand{\la}{\langle}
\newcommand{\ra}{\rangle}

\newlength{\dinwidth}
\newlength{\dinmargin}
\begin{document}  
\newcommand{\pom}{{I\!\!P}}
\newcommand{\reg}{{I\!\!R}}
\newcommand{\slowpi}{\pi_{\mathit{slow}}}
\newcommand{\fiidiii}{F_2^{D(3)}}
\newcommand{\fiidiiiarg}{\fiidiii\,(\beta,\,Q^2,\,x)}
\newcommand{\n}{1.19\pm 0.06 (stat.) \pm0.07 (syst.)}
\newcommand{\nz}{1.30\pm 0.08 (stat.)^{+0.08}_{-0.14} (syst.)}
\newcommand{\fiidiiiful}{F_2^{D(4)}\,(\beta,\,Q^2,\,x,\,t)}
\newcommand{\fiipom}{\tilde F_2^D}
\newcommand{\ALPHA}{1.10\pm0.03 (stat.) \pm0.04 (syst.)}
\newcommand{\ALPHAZ}{1.15\pm0.04 (stat.)^{+0.04}_{-0.07} (syst.)}
\newcommand{\fiipomarg}{\fiipom\,(\beta,\,Q^2)}
\newcommand{\pomflux}{f_{\pom / p}}
\newcommand{\nxpom}{1.19\pm 0.06 (stat.) \pm0.07 (syst.)}
\newcommand {\gapprox}
   {\raisebox{-0.7ex}{$\stackrel {\textstyle>}{\sim}$}}
\newcommand {\lapprox}
   {\raisebox{-0.7ex}{$\stackrel {\textstyle<}{\sim}$}}
\def\gsim{\,\lower.25ex\hbox{$\scriptstyle\sim$}\kern-1.30ex%
\raise 0.55ex\hbox{$\scriptstyle >$}\,}
\def\lsim{\,\lower.25ex\hbox{$\scriptstyle\sim$}\kern-1.30ex%
\raise 0.55ex\hbox{$\scriptstyle <$}\,}
\newcommand{\pomfluxarg}{f_{\pom / p}\,(x_\pom)}
\newcommand{\dsf}{\mbox{$F_2^{D(3)}$}}
\newcommand{\dsfva}{\mbox{$F_2^{D(3)}(\beta,Q^2,x_{I\!\!P})$}}
\newcommand{\dsfvb}{\mbox{$F_2^{D(3)}(\beta,Q^2,x)$}}
\newcommand{\dsfpom}{$F_2^{I\!\!P}$}
\newcommand{\gap}{\stackrel{>}{\sim}}
\newcommand{\lap}{\stackrel{<}{\sim}}
\newcommand{\fem}{$F_2^{em}$}
\newcommand{\tsnmp}{$\tilde{\sigma}_{NC}(e^{\mp})$}
\newcommand{\tsnm}{$\tilde{\sigma}_{NC}(e^-)$}
\newcommand{\tsnp}{$\tilde{\sigma}_{NC}(e^+)$}
\newcommand{\st}{$\star$}
\newcommand{\sst}{$\star \star$}
\newcommand{\ssst}{$\star \star \star$}
\newcommand{\sssst}{$\star \star \star \star$}
\newcommand{\tw}{\theta_W}
\newcommand{\sw}{\sin{\theta_W}}
\newcommand{\cw}{\cos{\theta_W}}
\newcommand{\sww}{\sin^2{\theta_W}}
\newcommand{\cww}{\cos^2{\theta_W}}
\newcommand{\trm}{m_{\perp}}
\newcommand{\trp}{p_{\perp}}
\newcommand{\trmm}{m_{\perp}^2}
\newcommand{\trpp}{p_{\perp}^2}
\newcommand{\alp}{\alpha_s}
\newcommand{\etamax}{\eta_{\rm max}}

\newcommand{\alps}{\alpha_s}
\newcommand{\sqrts}{$\sqrt{s}$}
\newcommand{\LO}{$O(\alpha_s^0)$}
\newcommand{\Oa}{$O(\alpha_s)$}
\newcommand{\Oaa}{$O(\alpha_s^2)$}
\newcommand{\PT}{p_{\perp}}
\newcommand{\JPSI}{J/\psi}
\newcommand{\sh}{\hat{s}}
\newcommand{\uh}{\hat{u}}
\newcommand{\MP}{m_{J/\psi}}
\newcommand{\PO}{I\!\!P}
\newcommand{\xbj}{x}
\newcommand{\xpom}{x_{\PO}}
\newcommand{\ttbs}{\char'134}
\newcommand{\xpomlo}{3\times10^{-4}}  
\newcommand{\xpomup}{0.05}  
\newcommand{\dgr}{^\circ}
\newcommand{\pbarnt}{\,\mbox{{\rm pb$^{-1}$}}}
\newcommand{\gev}{\,\mbox{GeV}}
\newcommand{\WBoson}{\mbox{$W$}}
\newcommand{\fbarn}{\,\mbox{{\rm fb}}}
\newcommand{\fbarnt}{\,\mbox{{\rm fb$^{-1}$}}}
\newcommand{\dsdx}[1]{$d\sigma\!/\!d #1\,$}
\newcommand{\eV}{\mbox{e\hspace{-0.08em}V}}
%
%
\newcommand{\qsq}{\ensuremath{Q^2} }
\newcommand{\gevsq}{\ensuremath{\mathrm{GeV}^2} }
\newcommand{\et}{\ensuremath{E_t^*} }
\newcommand{\rap}{\ensuremath{\eta^*} }
\newcommand{\gp}{\ensuremath{\gamma^*}p }
\newcommand{\dsiget}{\ensuremath{{\rm d}\sigma_{ep}/{\rm d}E_t^*} }
\newcommand{\dsigrap}{\ensuremath{{\rm d}\sigma_{ep}/{\rm d}\eta^*} }

\newcommand{\dstar}{\ensuremath{D^*}}
\newcommand{\dstarp}{\ensuremath{D^{*+}}}
\newcommand{\dstarm}{\ensuremath{D^{*-}}}
\newcommand{\dstarpm}{\ensuremath{D^{*\pm}}}
\newcommand{\zDs}{\ensuremath{z(\dstar )}}
\newcommand{\Wgp}{\ensuremath{W_{\gamma p}}}
\newcommand{\ptds}{\ensuremath{p_t(\dstar )}}
\newcommand{\etads}{\ensuremath{\eta(\dstar )}}
\newcommand{\ptj}{\ensuremath{p_t(\mbox{jet})}}
\newcommand{\ptjn}[1]{\ensuremath{p_t(\mbox{jet$_{#1}$})}}
\newcommand{\etaj}{\ensuremath{\eta(\mbox{jet})}}
\newcommand{\detadsj}{\ensuremath{\eta(\dstar )\, \mbox{-}\, \etaj}}

\def\Journal#1#2#3#4{{#1} {\bf #2} (#3) #4}
\def\NCA{\em Nuovo Cimento}
\def\NIM{\em Nucl. Instrum. Methods}
\def\NIMA{{\em Nucl. Instrum. Methods} {\bf A}}
\def\NPB{{\em Nucl. Phys.}   {\bf B}}
\def\PLB{{\em Phys. Lett.}   {\bf B}}
\def\PRL{\em Phys. Rev. Lett.}
\def\PRD{{\em Phys. Rev.}    {\bf D}}
\def\ZPC{{\em Z. Phys.}      {\bf C}}
\def\EJC{{\em Eur. Phys. J.} {\bf C}}
\def\CPC{\em Comp. Phys. Commun.}

\begin{titlepage}

\noindent

\noindent
\noindent

\vspace{2cm}
\begin{center}
\begin{Large}

{\bf Intermittency in Quantitative Finance
}
\end{Large}

\vspace{2cm}

Laurent Schoeffel \\~\\
CEA Saclay, Irfu/SPP, 91191 Gif/Yvette Cedex, \\
France
\end{center}

\vspace{2cm}

\begin{abstract}

Factorial
moments
are convenient tools  in nuclear physics to characterize the multiplicity
distributions when phase-space resolution ($\Delta$) becomes small.
For uncorrelated particle production within $\Delta$, 
Gaussian  statistics
holds  and factorial moments $F_q$ are equal to unity for all orders $q$. Correlations
between particles lead to a broadening of the multiplicity distribution 
and to dynamical fluctuations. In this case, the  factorial 
moments increase  above $1$
with decreasing $\Delta$. This corresponds to what can be called
intermittency.
In this letter, we show that a similar analysis can be developed
on financial price series, with an adequate definition of 
factorial moments.
An intermittent behavior can be extracted using moments of order $2$ ($F_2$),
illustrating a sensitivity to non-Gaussian fluctuations within
time resolution below $4$ hours.
This confirms that correlations
between price returns start to play a role when the time resolution
is below this threshold. 

\end{abstract}

\vspace{1.5cm}

\begin{center}
\end{center}

\end{titlepage}

%
%
%
%

\section{Introduction}

In this letter, we intend to show that  returns of
financial price series can be analyzed using standard techniques
of nuclear physics. We are interested in the multiplicity of
positive or negative returns in a given time window $\Delta$.
Indeed, sequences of positive and negative returns are much indicative of the
statistical nature of fluctuations in the price series.
The idea is then to extract a quantitative information from these
sequences. First, in this section, we present the  situation in
nuclear physics, that will be extended to finance in a second part.

At nuclear or sub-nuclear energies,
the number of hadrons created during inelastic
collisions varies from one event to another. 
The distribution of the number of produced hadrons, namely
the multiplicity distribution, 
provides a basic means to characterize the events.
The multiplicity distribution contains information 
about multi-particle correlations in an integrated form,
providing a general and sensitive means to probe
the dynamics of the interaction.
Particle multiplicities  
can be studied in terms of the 
normalized factorial moments
\begin{equation}
\label{facm}
F_q(\Delta)=\la n(n-1)\ldots (n-q+1)\ra / \la n \ra^q,
\qquad q=2,3, \ldots ,  
\end{equation}
for a specified phase-space region of size $\Delta$. 
The number, $n$, of particles is measured inside  $\Delta$
and angled brackets $\la\ldots\ra$ denote 
the average over all events. The factorial
moments
are convenient tools  to characterize the multiplicity
distributions when $\Delta$ becomes small.
For uncorrelated particle production within $\Delta$, 
Poisson or Gaussian  statistics
holds  and $F_q=1$ for all $q$. Correlations
between particles lead to a broadening of the multiplicity distribution 
and to dynamical fluctuations. In this case, the normalized factorial 
moments increase  
with decreasing $\Delta$. 
The idea is then to divide the factorial moment defined in Eq. (\ref{facm})
in more and more bins. 

We can thus compute the related moment following
Eq. (\ref{facm}) as
\be
F_{i}=\frac{1}{N_{events}}\sum_{events}
\frac{\sum_{k=1}^{n_{bins}} \left\{n_{k}(n_{k}-1)
\cdots (n_{k}-i+1)\right\}/n_{bins}}
{(\langle n\rangle /n_{bins})^{i}}
\label{genfacmom}
\ee
where $\langle n \rangle$ is the average number of particles in the full
phase space region accepted ($\Delta$), $n_{bins}$ denotes the number of bins in
this region and $n_k$ is the multiplicity in $k$-th bin.

The behavior of factorial moments plotted as a function of the number of bins 
(which means decreasing bin sizes)
provides information about the character of multiplicity
fluctuations among different bins. Rising of $F_i$ with rising $n_{bins}$
(decreasing bin size) generally signalizes deviation from purely
Gaussian distribution of fluctuations. The linear growth
of $\log F_i$ with $n_{bins}$ was defined as intermittency in \cite{BiaP}. 
See also \cite{Lipa,Dremin:1993dt,Dremin:1993ee,Rames:1994qm}. In the following, 
the term is  used for any
type of growth of $F_i$ observed.

\section{Application to financial price series}

The analogy with returns of financial price series is immediate.
If we divide the price series $y(t)$ in consecutive time windows of lengths $\Delta$,
we define like this a set of events.
In each window, we have a certain number of positive returns  $n_+$,
where $y(t)-y(t-1) >0$, and similarly of
negative returns $n_-$.
If the sequence of returns is purely
 uncorrelated, following a  
Gaussian  statistics at all scales,
we expect $F_q=1$ for all $q$. 

However, correlations
between returns may lead to a broadening of the multiplicity distributions 
($n_+$ or $n_-$ or even a combination of both) 
and to dynamical fluctuations. In this case, the  factorial 
moments may increase  
with decreasing $\Delta$, or increasing the number of bins that divide $\Delta$, as in
Eq. (\ref{genfacmom}). In the following, we consider only the factorial moment of
second order $F_2$.
We can write
\be
F_{2}^{++}=\frac{1}{N_{events}}\sum_{events}
\frac{\sum_{k=1}^{n_{bins}} \left\{n_{k}^{+}(n_{k}^{+}-1)
\right\}/n_{bins}}
{(\langle n^{+}\rangle /n_{bins})^{2}}
\label{like}
\ee
where $\langle n \rangle$ is the average number of positive returns in the full
time window  ($\Delta$), $n_{bins}$ denotes the number of bins in
this window and $n_k^{+}$ is the number of positive returns in $k$-th bin.

\begin{figure}[htbp]
  \begin{center}
    \includegraphics[width=0.8\textwidth]{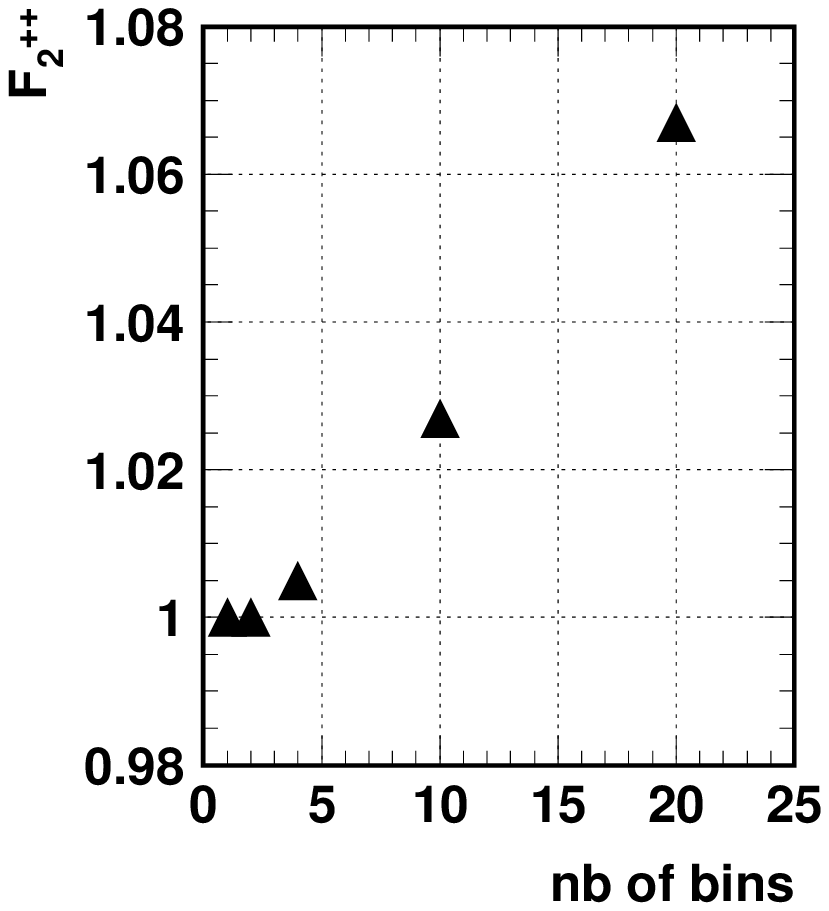}
  \end{center}
  \caption{Euro future contract. Factorial moments $F_2^{++}$ are displayed for $1$, $2$, $4$, $10$ or $20$ bins.
The global time window of $16.5$ hours ($n_{bins}=1$).
This provides a time resolution ranging from $1.6$ hours for $20$ bins till $16.5$ hours for $1$ bin.
We observe that for $1$ and $2$ bins segmentation, $F_2^{++}$ is found equal to $1$
and $F_2^{++}$ is increasing above $1$ when the number of bins gets larger than $4$.
This confirms that non-Gaussian fluctuations in the sequence of returns
returns start to play a role when the resolution
is below $4$ hours. }
\label{f2pp}
\end{figure}

We consider the data  series on  the Euro future contract, sampled in 5 minutes units
from may 2001 till August 2011, which counts 688k quotes.
We present calculations for a time window 200 times units, that we
divide in $1$, $2$, $4$, $10$ or $20$ bins.
This means that the  time resolution extends from $1.6$ hours to  $16.5$ hours.
Note that with a time window of 200 time units, we set up an ensemble of more that 3400 events.
The statistical precision of the following analysis is then ensured.
Results are presented in Fig. \ref{f2pp} for positive returns.
Factorial moments $F_2^{++}$ are displayed for $1$, $2$, $4$, $10$ or $20$ bins.
As mentioned above, this gives a time resolution ranging from $1.6$ hours for $20$ bins till $16.5$ hours for $1$ bin.
The statistical precision is of $0.1$\%. We can define a systematical uncertainty
by shifting the time window of $200$ units by $50$ or $100$ units, which means that 
we define a different set of events among the price series. Variations in the
calculations of $F_2^{++}$ are lower that $0.1$\%. Fig. \ref{f2pp} displays the
full uncertainty of these quantities.

In Fig. \ref{f2pp},
we observe that for $1$ and $2$ bins segmentation $F_2^{++}$ is found to be equal to $1$.
As expected, for the larger resolution, positive returns appear as completely uncorrelated.
When the number of bins is increased, we observe the phenomenon described in 
section 1, with an enhanced sensitivity of $F_2^{++}$ to non-Gaussian fluctuations.
This confirms that correlations
between positive returns start to play a role when the resolution
is below $4$ hours. 
Thus, Fig. \ref{f2pp} exhibits a clear feature of intermittency.
As in nuclear reactions, an increase of the resolution
to a certain extend leads to a broadening of the multiplicity distribution 
and to super-diffusive fluctuations. Let us note that this is a feature that can be
approached in the context of non-extensive statistics \cite{ts1,ts2,ts3}.

Similar results can be obtained for $F_2^{--}$, defined for negative returns distribution.
\be
F_{2}^{--}=\frac{1}{N_{events}}\sum_{events}
\frac{\sum_{k=1}^{n_{bins}} \left\{n_{k}^{-}(n_{k}^{-}-1)
\right\}/n_{bins}}
{(\langle n^{-}\rangle /n_{bins})^{2}}
\label{like}
\ee
where $\langle n \rangle$ is the average number of negative returns in the full
time window  ($\Delta$), $n_{bins}$ denotes the number of bins in
this window and $n_k^{-}$ is the number of negative returns in $k$-th bin.
For all values displayed in Fig. \ref{f2pp} for $F_2^{++}$, we derive the
same result for $F_2^{--}$ 
 up to $0.1$\%, which makes $F_2^{++}$ and $F_2^{--}$ 
indistinguishable. 

\begin{figure}[htbp]
  \begin{center}
    \includegraphics[width=0.8\textwidth]{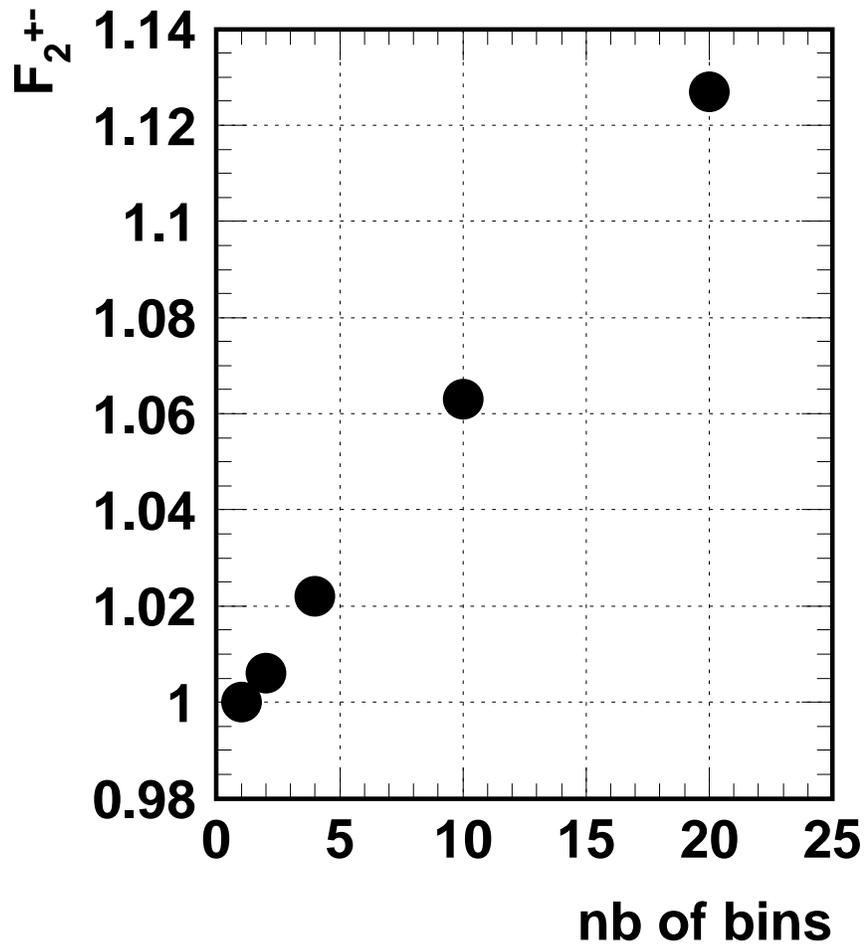}
  \end{center}
  \caption{Euro future contract. Factorial moments $F_2^{+-}$ are displayed for $1$, $2$, $4$, $10$ or $20$ bins.
The global time window of $16.5$ hours ($n_{bins}=1$).
This provides a time resolution ranging from $1.6$ hours for $20$ bins till $16.5$ hours for $1$ bin.
We observe that for $1$ bin, $F_2^{+-}$ is found equal to $1$
and $F_2^{+-}$ is increasing above $1$ when the number of bins gets larger than $2$.
Intermittency is enhanced compared to the like sign quantity $F_2^{++}$. }
\label{f2pm}
\end{figure}

\begin{figure}[htbp]
  \begin{center}
    \includegraphics[width=0.8\textwidth]{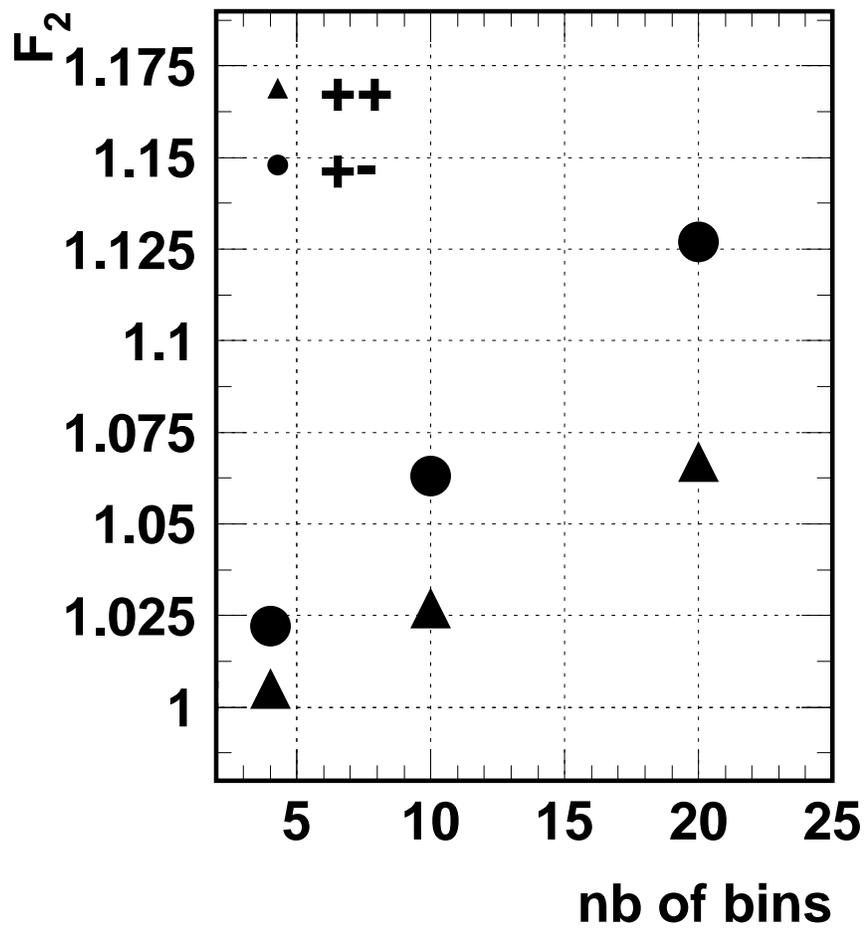}
  \end{center}
  \caption{Euro future contract. Factorial moments $F_2^{++}$ and $F_2^{+-}$ are displayed for $1$, $2$, $4$, $10$ or $20$ bins.
The global time window of $16.5$ hours ($n_{bins}=1$). See text for details.}
\label{f2comb}
\end{figure}

A direct extension of the above study can be
obtained if we examine moments for like-sign and
unlike-sign combinations of returns separately.
The like-sign factorial moment of order $2$ is defined by Eq. (\ref{like}).
The unlike-sign can be expressed as 
\be
F_{2}^{+-}=\frac{1}{N_{events}}\sum_{events}
\frac{\sum_{k=1}^{n_{bins}} \left\{n_{k}^{+}n_{k}^{-}
\right\}/n_{bins}}
{\langle n^{+}\rangle \langle n^{-}\rangle /(n_{bins})^{2}}
\label{unlike}
\ee
Results are presented in Fig. \ref{f2pm} for $F_{2}^{+-}$.
Here again, we observe intermittency, with an increase of $F_{2}^{+-}$ 
as a function of the number of bins.
Also, this increase is larger than for like-sign calculations.
In order to illustrate this point more clearly,
a combination of values from Fig. \ref{f2pp} and \ref{f2pm}
is presented in Fig. \ref{f2comb} for a number of bins larger than $4$.
The sensitivity to non-Guassian fluctuations in the returns sequence 
is thus enhanced with the definition of d to $F_{2}^{+-}$. This is also an effect
observed in nuclear interactions \cite{Dremin:1993dt,Dremin:1993ee,Rames:1994qm}.

\begin{figure}[htbp]
  \begin{center}
    \includegraphics[width=0.8\textwidth]{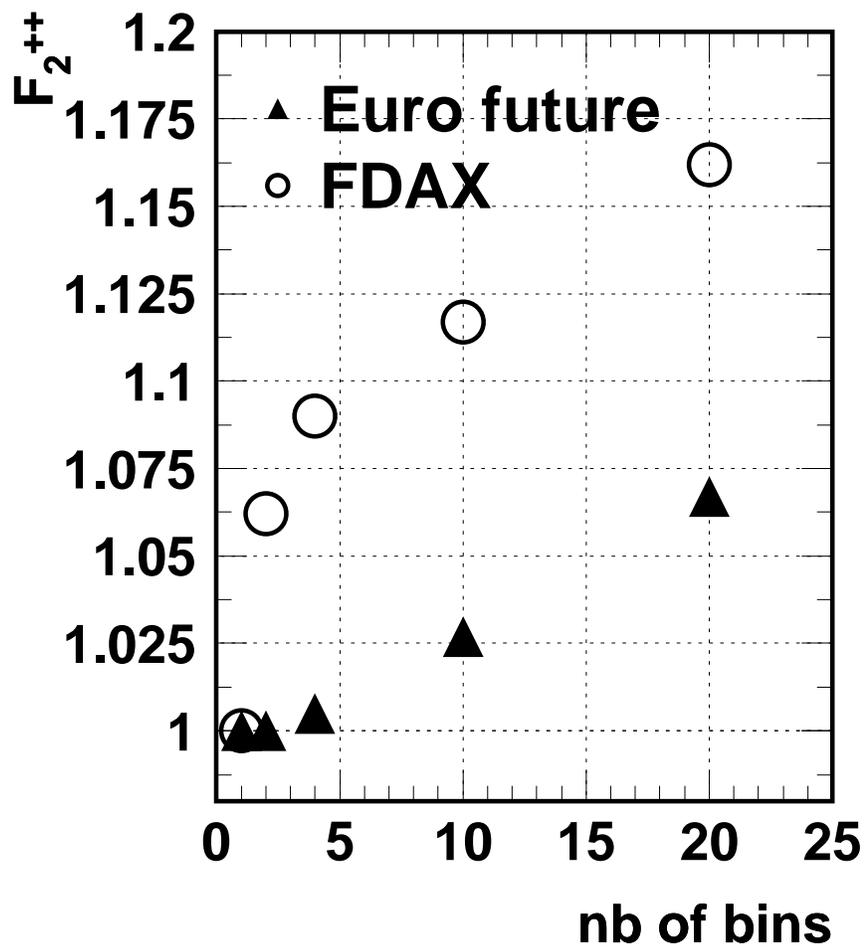}
  \end{center}
  \caption{Euro and DAX future contracts. Factorial moments $F_2^{++}$ are displayed for $1$, $2$, $4$, $10$ or $20$ bins.
The global time window of $16.5$ hours ($n_{bins}=1$).
This provides a time resolution ranging from $1.6$ hours for $20$ bins till $16.5$ hours for $1$ bin.
For the DAX future, the intermittent behavior is more pronounced than for the Euro. }
\label{f2pp2}
\end{figure}

A final comment is in order. This analysis has been illustrated on the Euro future contract.
However,
we have found that a similar intermittent behavior is observed on other futures, 
like the DAX future (FDAX) and the Pound future and commodities.
For each series, the values of the like-sign and unlike-sign factorial moments of order $2$
vary compared to results exhibited for the Euro future.
But the rise of the $F_2$ as a function of the number of bins is an invariant
property, with always $F_2^{++}$ and $F_2^{+-}$ equal to unity for a time window of
order $16$ hours. For some series, like FDAX, the intermittent behavior already prevails for
a time resolution of order $8$ hours, with an increase of $F_2^{++}$ from $1$ ($n_{bins}=1$, $\delta T\sim16$~h)
to $1.16$ ($n_{bins}=20$, $\delta T\sim1.6$~h).
This is shown in Fig. \ref{f2pp2}.
This growth with $n_{bins}$ is faster than the Euro future. However, the key feature is always the same.
For a sufficiently fine resolution in time, more precisely for a resolution $\delta T$ smaller than $4$ to $8$ hours
depending of the financial product, intermittency  is observed.

\section{Conclusion}

Factorial
moments
are convenient tools  in nuclear physics to characterize the multiplicity
distributions when phase-space resolution becomes small.
In particular,
correlations
between particles lead to a broadening of the multiplicity distribution 
and to dynamical fluctuations. In this case, the  factorial 
moments increase  above $1$
with decreasing resolution. This corresponds to what can be called
intermittency.

A similar property has been illustrated on financial price series,
by extending the concepts from nuclear physics to quantitative finance.
An intermittent behavior has been extracted 
for different price series on future contracts,
using moments of order $2$ ($F_2$). This
illustrates a sensitivity to non-Gaussian fluctuations within
time resolution smaller than $4$ to $8$ hours.
This confirms that correlations
between price returns start to play a role when the time resolution
is below a threshold of a few hours. 

Clearly, this technique can be used as a global classification of financial
products as a function of the strength of the intermittency. 
This can be seen as a prerequisite to more involved analysis of these time series \cite{ls}.


\newpage

\end{document}